\documentclass[11pt]{article}

\usepackage{color,graphicx}
\usepackage{young}
\usepackage[vcentermath]{youngtab}
\usepackage{amsmath,amssymb,graphicx}
\usepackage{hyperref}
\definecolor{darkred}{rgb}{0.65,0.15,0}
\hypersetup{pdfborder={0 0 0},colorlinks=true,urlcolor=darkred,citecolor=blue,linkcolor=darkred,linktocpage=true}

\usepackage{cite}
\usepackage{amsmath}
\usepackage{amsfonts}
\usepackage{amssymb}
\usepackage{graphicx}%
\usepackage{amsthm}
\usepackage{mathrsfs}
\usepackage[T1]{fontenc}
\usepackage{enumerate}
\usepackage{color}
\usepackage{hyperref}
\hypersetup{
   colorlinks   =  true,
    citecolor    = red,
     urlcolor	=magenta,
}

\def\4diml{four-dimensional}

\def\-1{^{-1}}

\newcommand{\G}{\mathscr{G}}

\makeatletter

\@addtoreset{equation}{section}
\makeatother

\begin{document}

\thispagestyle{empty}

\vspace{5mm}

\begin{center}
{\LARGE \bf  Integrable sigma models with Haantjes structure\\[2mm] on the ${H_{4}}$ Lie group}

\vspace{14mm}
\normalsize
{\large Mirenayatollah Bahadori \footnote{en.bahadori@gmail.com }, Ali Eghbali\footnote{eghbali978@gmail.com}, Adel Rezaei-Aghdam\footnote{Corresponding author: rezaei-a@azaruniv.ac.ir}}

\vspace{4mm}
{\small {\em Department of Physics, Faculty of Basic Sciences,\\
Azarbaijan Shahid Madani University, 53714-161, Tabriz, Iran}}\\

\begin{abstract}
 By solving algebraic relations for the conditions of Haantjes structure on a Lie algebra ${\G}$ and by using the corresponding automorphism group we proceed
 to classify all inequivalent algebraic Haantjes structures on ${\G}$. In this manner,
 we obtain 49 inequivalent algebraic Haantjes structures on the ${h_{4}}$ Lie algebra.
We propose a new deformation of the chiral sigma model on a Lie group by using Haantjes structure on it.
Then, we will try to obtain conditions on this structure such that the deformed sigma model remains to be integrable.
Finally, using the ${h_{4}}$ Haantjes structures and solving the conditions on the integrability of the model,
we find three new integrable sigma models on the ${H_{4}}$ Lie group.
 \end{abstract}
\end{center}
{ $~~~~~~$ {\bf Keywords:} Lie group, Haantjes structure, Integrable sigma model}

\newpage
\setcounter{page}{1}

\tableofcontents

 \vspace{5mm}
 \vspace{5mm}

\section{\label{Sec.II} Introduction}
Two-dimensional integrable sigma models and their deformations have attended considerably over the past 45 years \cite{KP, VG, HE1, HE2}.
The integrable deformation of the principal chiral model on $SU(2)$ was first considered in \cite{BP, CI, VA, NJ}.
Klimcik introduced \cite{CK} the Yang-Baxter sigma model as a particular type of deformation of the principal chiral model which was built on
an arbitrary Lie group. Then, the integrability of the model was proven in \cite{CKL} (see, also, \cite{CKL1,Yoshida.cybe}).
The name originates from the anti-symmetric classical r-matrix solving the inhomogeneous
Yang-Baxter deformation used to construct the action of the model.
The Yang-Baxter type deformation was then generalized by Delduc, Magro and Vicedo to the symmetric space sigma model in \cite{Delduc1}
and to the $AdS_5 \times S^5$ superstring based on the solutions of the modified classical Yang-Baxter equation in \cite{Delduc2}
and classical Yang-Baxter equation in \cite{Kawaguchi1}, resulting in a quantum deformation of the superconformal symmetry of $AdS_5 \times S^5$ \cite{Delduc3}.
This type of deformation was later rapidly developed by many authors in the superstring backgrounds \cite{superstring.back}.
The generalization to Yang-Baxter sigma models with WZW term was also carried out in \cite{Delduc4},
and thus the Yang-Baxter Wess-Zumino-Witten model was introduced together with its integrability properties.
On can find various aspects of the Yang-Baxter Wess-Zumino-Witten models along with some examples in
\cite{Yoshida.NPB,EP, EP1}.
Recently, it has been introduced an integrable sigma model on Lie groups equipped with a generalized complex structure and generalized $\mathcal{F}$ structure \cite{RA, EN}. In all of the above models one uses the Nijenhuis operator with zero torsion \cite{JP}.
When the Nijenhuis torsion is not zero one can use Haantjes operator which presented by Haantjes \cite{JP1}.
Recently, this structure has been used in studying the dynamical systems and their integrability \cite{JP2,JP3,JP4}.
Here we will try to use the Haantjes structure to build some integrable sigma models.
First, we write the formulation of the Haantjes structure in the Lie algebra framework. Then,
we solve the algebraic relations for the conditions of
Haantjes structure, and by using the automorphism group of the ${h_{4}}$ Lie algebra
we obtain all the corresponding inequivalent algebraic Haantjes structures.
Most importantly, we propose a new deformation of the chiral sigma model on a Lie group by using Haantjes structure on it.
Finally, using the ${h_{4}}$ Haantjes structures and solving the conditions on the integrability of the model,
we find three new integrable sigma models on the ${H_{4}}$ Lie group.
Recently, the Yang-Baxter deformations of Wess-Zumino-Witten model on the ${H_{4}}$ Lie group have been examined in \cite{EP}.
Then, by applying the Poisson-Lie T-duality transformations \cite{Klim1,Klim2}, it has been obtained the non-Abelian target space duals of
those deformed models \cite{egh.gh.rez}.
In the present work, by comparing our results with deformed backgrounds of the ${H_{4}}$ and their non-Abelian duals,
we show that our results are different from those of \cite{EP} and \cite{egh.gh.rez}.

The plan of the paper is outlined as follows:
In Section 2 after reviewing some concepts about Haantjes structure we obtain algebraic relations for the Haantjes structure.
Then, in Section 3 by solving the algebraic (matrix) relations for the ${h_{4}}$ Lie algebra and then by using its corresponding automorphism group we classify the ${h_{4}}$ Haantjes structures into 49 inequivalent families.
After then in Section 4 we deform the chiral sigma model on a Lie group by using the Haantjes structure on it.
We obtain the conditions on Haantjes structure such that the deformed sigma model remains to be integrable.
Finally, in Section 5, by using Haantjes structures on the ${h_{4}}$ and the integrability conditions on the deformed chiral sigma model we obtain three new integrable sigma models on the ${H_{4}}$ Lie group.

\section{\label{Sec.II} A review of Haantjes structure }

Let $M$ be an m-dimensional real differentiable manifold and $\boldsymbol{L}:TM\rightarrow TM$ be a smooth $(1,1)$ tensor field. The
 \textit{Nijenhuis torsion} of $\boldsymbol{L}$ is  defined by the vector-valued $2$-form
\begin{equation}\label{2.1}
\forall~~  X,Y \in  TM~~~N(X,Y):= \boldsymbol{L}^2[X,Y] +[\boldsymbol{L}X,\boldsymbol{L}Y]-\boldsymbol{L}\Bigl([X,\boldsymbol{L}Y]+[\boldsymbol{L}X,Y]\Bigr),
\end{equation}
where  $[ \ , \ ]$ denotes the commutator of two vector fields.
In local coordinates $\boldsymbol{x}=(x^1,\ldots, x^m)$ we have $\boldsymbol{L}=L^\mu_\nu dx^{\nu}\otimes \partial_{\mu}$ and the Nijenhuis torsion $N=N^{\mu}_{\nu\lambda}\partial_{\mu}\otimes dx^{\nu}\otimes dx^{\lambda}$, then the components of $N$ can be written as \cite{JP3, JP4}
\begin{equation}
N^\mu_{~\nu\lambda}=\sum_{\rho=1}^m\biggl(\partial_{\rho}{\boldsymbol{L}}^\mu_\lambda{\boldsymbol{L}}^\rho_\nu -\partial_{\rho} {\boldsymbol{L}}^\mu_\nu {\boldsymbol{L}}^\rho_\lambda+\Bigl(\partial_{\lambda} {\boldsymbol{L}}^\rho_\nu  -\partial_{\nu} {\boldsymbol{L}}^\rho_\lambda \Bigr) {\boldsymbol{L}}^\mu_\rho \biggr)\ ,
\end{equation}
with  $m^2(m-1)/2$ independent components.
If the Nijenhuis torsion is zero then $\boldsymbol{L}$ is called Nijenhuis operator. For the case of $\boldsymbol{L}^{2}=-1$ and even $m$, $\boldsymbol{L}$ is called complex structure. When $N\neq0$ then
 the following \textit{Haantjes torsion} $\mathcal{H}_{\boldsymbol{L}}$ can be associated to $\boldsymbol{L}$ such that it is a vector-valued $2$-form defined by \cite{JP1, JP2, JP3, JP4}
\begin{equation} \label{2.3}
\forall~~  X,Y \in  TM~~~\mathcal{H}_{\boldsymbol{L}}(X,Y) := \boldsymbol{L}^2N(X,Y)+N(\boldsymbol{L}X,\boldsymbol{L}Y)-\boldsymbol{L}\Bigl(N(X,\boldsymbol{L}Y)+N(\boldsymbol{L}X,Y)\Bigr).
\end{equation}
One can also write the above relation explicitly as \cite{JP2}
\begin{eqnarray*}\label{2.4}
\mathcal{H}_{\boldsymbol{L}}(X,Y)&=&\boldsymbol{L}^4 [X,Y] -2\boldsymbol{L}^3\Bigl([X,\boldsymbol{L}Y]+[\boldsymbol{L}X , Y]\Bigr)
+\boldsymbol{L}^2\Bigl( [X, \boldsymbol{L}^2 Y]+4\, [\boldsymbol{L}X,\boldsymbol{L}Y]+[\boldsymbol{L}^2X,Y]\Bigr)
\end{eqnarray*}
\begin{equation}
-2 \boldsymbol{L}\Bigl([\boldsymbol{L}X,\boldsymbol{L}^2Y]+[\boldsymbol{L}^2X,\boldsymbol{L}Y]
\Bigr)+[\boldsymbol{L}^2X,\boldsymbol{L}^2Y]
\ .
\end{equation}
In the local coordinates $x^{\mu}$ the components of the Haantjes torsion can be written as follow \cite{JP3, JP4}.
\begin{equation}
(\mathcal{H}_{\boldsymbol{L}})^\mu_{\nu\lambda}=  \sum_{\alpha,\beta=1}^n\biggl(
\boldsymbol{L}^\mu_\alpha \boldsymbol{L}^\alpha_\beta(N)^\beta_{\nu\lambda}  +
(N)^\mu_{\alpha \beta}\boldsymbol{L}^\alpha_\nu \boldsymbol{L}^\beta_\lambda-
\boldsymbol{L}^\mu_\alpha\Bigl ((N)^\alpha_{\beta \lambda} \boldsymbol{L}^\beta_\nu+
 (N)^\alpha_{\nu \beta } \boldsymbol{L}^\beta_\lambda \Bigr)
 \biggr) \ .
 \end{equation}
When the Haantjes torsion for $\boldsymbol{L}$ is zero then $\boldsymbol{L}$ is called Haantjes operator or structure.

\subsection{\label{subSec.II.1} Haantjes structure on a Lie algebra (group)}
Suppose that manifold $M$ is a Lie group $G$ with Lie algebra ${\G}$ defined by the bases $\{T_i\} ,i=1, ..., m,$ where
\begin{eqnarray}\label{2.6}
[T_i ,  T_j] = {f}^{k}_{_{~ij}}~T_k ,
\end{eqnarray}
in which ${f}^{k}_{_{~ij}}$ are the structure constants of ${\G}$. One can write Haantjes operator $\boldsymbol{L}$ in the local coordinates $x^{\mu}$ of the group manifold in the following form \cite{nappi1993wess}
\begin{eqnarray}\label{2.5}
\boldsymbol{L}=e^{\mu}_{~i}e_{\nu}^{~j}L^{i}_{~j}\partial_{\mu}\otimes dx^{\nu} ,
\end{eqnarray}
where $e^{\mu}_{~i}$ are vielbeins for the Lie group $G$ with their inverse $e_{\nu}^{~j}$ ( $e^{\mu}_{~i}e_{\nu}^{~j}=\delta^{\mu}_{\nu}$) \cite{nappi1993wess}. Now, assume that $L^{i}_{~j}$'s are the coefficients of the map     $\cal{L}:\G\rightarrow \G$ such that
\begin{eqnarray}
 {\cal L}(T_{i})={L}^j_{~i}~T_{j}.
\end{eqnarray}
Note that $L^{i}_{~j}$'s are the independent of the Lie group coordinates. In this meaner one can rewrite relation \eqref{2.4} in the following form for the components of Haantjes torsion in term of components $L^{i}_{~j}$:
\begin{eqnarray}\label{2.7}
&&(\mathcal{H}_{\boldsymbol{L}})^p_{~ij}:
={f}^k_{~ij}~{L}^l_{~k}~{L}^m_{~~l}~{L}^n_{~m}~{L}^p_{~n}-2{L}^k_{~j}~{f}^l_{~ik}~{L}^m_{~~l}~{L}^n_{~m}~{L}^p_{~n}
-2{L}^k_{~i}~{f}^l_{~kj}~{L}^m_{~~l}~{L}^n_{~m}~{L}^p_{~n}\nonumber\\
&&~~~~~~~~~~~~~+{L}^k_{~j}~{L}^l_{~k}~{f}^m_{~il}~{L}^n_{~m}~{L}^p_{~n}
+4{L}^k_{~i}~{L}^l_{~j}~{f}^m_{~kl}~{L}^n_{~m}~{L}^p_{~n}+{L}^k_{~i}~{L}^l_{~k}~{f}^m_{~lj}~{L}^n_{~m}~{L}^p_{~n}\nonumber\\
&&~~~~~~~~~~~~~-2{L}^k_{~i}~{L}^l_{~j}~{L}^m_{~~l}~{f}^n_{~km}~{L}^p_{~n}-2{L}^k_{~i}~{L}^l_{~k}~{L}^m_{~~j}~{f}^n_{~lm}~{L}^p_{~n}
+{L}^k_{~i}~{L}^l_{~k}~{L}^m_{~~j}~{L}^n_{~m}~{f}^p_{~ln}=0.~~~\label{2.7}
\end{eqnarray}
In calculating the above relation, we have used the following Maurer-Cartan equation \cite{nappi1993wess}
\begin{eqnarray}\label{mc}
f^{i}\hspace{0cm}_{jk}=e_{\mu}^{~i} (e^{\nu}_{~j}  \partial_{\nu} e^{\mu}_{~k}- e^{\nu}_{~k}  \partial_{\nu} e^{\mu}_{~j}).
\end{eqnarray}
It is also useful to obtain matrix form of \eqref{2.7} with the condition $(\mathcal{H}_{\boldsymbol{L}})^p_{~ij}=0$.
In this way, one may use the $(L)^{i}_{~j}=L^{i}_{~j}$ and the adjoint representation
$({\cal X}_{j})_{k}^{~i}=-f^{i}\hspace{0cm}_{jk}$ to get\footnote{Here ``t'' stands for the transposition.}
\begin{flalign}\label{2.11}
& \hspace{4cm} -{\cal X}_{_i}(L^{t})^{4} + 2L^{t} {\cal X}_{_i} (L^{t})^{3} + 2{L}^k_{~i} {\cal X}_{_k} (L^{t})^{3} & \nonumber \\
& \hspace{4cm} ~~~~~~~~~~~~~~~- (L^{t})^{2} {\cal X}_{_i} (L^{t})^{2} - 4{L}^k_{~i}L^{t} {\cal X}_{_k} (L^{t})^{2} & \nonumber \\
& \hspace{4cm}  ~~~~~~~~~~~~~~~- ((L^{t})^{2})_i^{~l} {\cal X}_{_l} (L^{t})^{2} + 2{L}^k_{~i}(L^{t})^{2} {\cal X}_{_k} L^{t} & \nonumber \\
& \hspace{4cm}  ~~~~~~~~~~~~~~~+ 2((L^{t})^{2})_i^{~l}L^{t} {\cal X}_{_l} L^{~t} - ((L^{t})^{2})_i^{~l}(L^{t})^{2} {\cal X}_{_l} = 0. &
\end{flalign}
We call $L^{i}_{~j}$ or matrix $L$ the Haantjes structure on Lie algebra $\G$.
In order to obtain the Haantjes structure on a Lie algebra (group) it is enough to solve the above matrix relation.
For obtaining inequivalent structures one must use Proposition given below,
where it is used the automorphism group $A:\G\rightarrow \G$ of the Lie algebra $\G$ with \eqref{2.6} such that
\begin{eqnarray}\label{2.12}
T'_{i}=A(T_{i})=A_{i}^{~j}T_{j}.
\end{eqnarray}
By inserting \eqref{2.12} into $[T'_i ,  T'_j] = f^{k}_{~ij} ~T'_k$ we obtain
\begin{eqnarray}\label{2.14}
{A}_i^{~m}~ f^{k}_{~mn}~ {A}_j^{~n} = f^{l}_{~ij}~ A_l^{\;\;k}.
\end{eqnarray}
Because of tensorial form of the above equation, working with it is not so easy and we suggest writing this equation as matrix
form using the adjoint representation $({\cal X}_{m})_{n}^{~k}=-f^{k}_{~mn}$. Then, one obtains
\begin{align}\label{2.14.1}
&~{A}_i^{~m}~ ({\cal X}_{m})_{n}^{~k}~ {A}_j^{~n} = ({\cal X}_{i})_{j}^{~l}~ A_l^{\;\;k}\nonumber\\
&\Rightarrow {A}_j^{~n} ~ ({A}_i^{~m} {\cal X}_{m})_{n}^{~k} = ({\cal X}_{i} A)_{j}^{~k}\nonumber\\
&\Rightarrow A ~ ({A}_i^{~m} {\cal X}_{m}) = {\cal X}_{i} A.
\end{align}
Multiplying both sides of the above equation in $A^{-1}$, we finalize that
\begin{eqnarray}\label{2.15}
{A}_i^{~m} {\cal X}_{m} = A^{-1} {\cal X}_{i} A.
\end{eqnarray}
\\
{\bf Proposition.} {\it Let $L$ and $L'$ be two Haantjes structures  as solutions of  \eqref{2.11}.
If there exists an automorphism $A$ of $\G$ such that
\begin{eqnarray}\label{2.17}
{L'}^j_{~i} = (A^{-1})_i^{~k}~{L}^l_{~k}A_l^{~j},
\end{eqnarray}
or
\begin{eqnarray}\label{2.16}
{L^\prime}^{t} = A^{-1}~L^{t} ~A,
\end{eqnarray}
then it is said the Haantjes structures $L$ and $L'$ on $\G$ are equivalent. }\\
\\
{\it Proof.}~Let $\{T_i\}$ and $\{T'_i\}$ be the bases of $\G$ such that $T'_i = {A}_i^{~j}~ T_j$ in which ${A}_i^{~j}$
is an element of automorphism group  $Aut (\G)$. Assume that relation \eqref{2.11} holds for ${L'}^j_{~i}$'s,
applying \eqref{2.16} in \eqref{2.11} and then using \eqref{2.15} we arrive at \eqref{2.11} for $L^{i}_{~j}$.

In the next section we will use the above Proposition to obtain the inequivalent Haantjes structures on the $h_{4}$.
Indeed, the structures relating to each other through \eqref{2.16} are equivalent.
In fact, one can use \eqref{2.16} to obtain inequivalent Haantjes structures.

\section{Haantjes structure on the $h_{4}$ Lie algebra}
Before  proceeding further let us the consider the automorphism group of the particular $h_{4}$ Lie algebra. This Lie algebra is generated by the set of generators $(T_{1}, T_{2}, T_{3}, T_{4})$ with the following commutation relations
rules:
\begin{eqnarray}\label{3.1}
[T_{1} , T_{2}]~=~T_{2},~~~~~[T_{1} ,T_{3}]~=~-T_{3},~~~~~[T_{3} , T_{2}]~=~T_{4},
\end{eqnarray}
where $T_{4}$ is a central generator. The elements of automorphism group of the $ h_{4} $ can be calculated as follows \cite{EP}
{\small \begin{eqnarray}
Aut(h_4)=\left\{ A_1=\left( \begin{tabular}{cccc}
                 1 & $c$ & $d$ & $e$ \\
                 0 & $a$ & 0 & $ad$ \\
                 0 & 0 & $b$ & $bc$ \\
                 0 & 0 & 0 & $ab$ \\
                 \end{tabular} \right),~A_2=\left( \begin{tabular}{cccc}
                 -1 & $c$ & $d$ & $e$ \\
                 0 & 0  & $a$ & -$ac$ \\
                 0 & $b$ & 0 & -$bd$ \\
                 0 & 0 & 0 & -$ab$ \\
                 \end{tabular} \right); ~ ab \neq 0\right \},\label{3.2}
\end{eqnarray}}
where $a, b, c, d$ are arbitrary real constants.
Now let us consider the general solutions of $4 \times 4$ matrix relations of \eqref{2.11} for the $ h_{4} $.
In this part we have used the computer program Maple 2024 and have shown that there are 64 solutions for these relations,
but only 49 of them are inequivalent\footnote{Note that equation \eqref{2.11} contains fourth powers of $L$ and it does not seem a priori obvious that one could easily classify all possible solutions. As mentioned in the above, we obtain the general solutions of
\eqref{2.11} for the $h_4$ using the computer program Maple.
Therefore, it cannot be definitively said that Maple has provided us with all possible general solutions.}.
For the sake of clarity the obtained results are summarized in Table 1;
we display the non-zero components of the inequivalent Haantjes structures on the $h_{4}$.
We denote each structure by $L^{(h_{4})}_{_{i}}$, where the index $i$ distinguishes between several possible
inequivalent structures.

As an example let us consider the following general solution
{\small \begin{eqnarray}\label{3.3}
	L~=~\left( \begin{tabular}{cccc}
	${L}^1_{~1}$ & $0$ & $0$ & $0$\\
	${L}^2_{~1}$ & ${L}^2_{~2}$ & $0$ & $0$ \\
	$0$ & $0$ & ${L}^3_{~3}$ & $0$ \\
	${L}^4_{~1}$ & $0$ & ${L}^4_{~3}$ & ${L}^4_{~4}$\\
	\end{tabular} \right),
\end{eqnarray}}
where ${L}^3_{~3} = {L}^4_{~4}$. Here we shall consider ${L}^1_{~1} =\alpha_{2}$, ${L}^2_{~2}=\beta_{2}$ and
${L}^3_{~3} = {L}^4_{~4} =\gamma_{2}$, where $\alpha_{2}$, $\beta_{2}$ and $\gamma_{2}$
are some real constant parameters.
Using \eqref{2.16}  together with the automorphism transformation \eqref{3.2} one can show that the solution \eqref{3.3} is equivalent to the following one
\begin{eqnarray}\label{3.3.1}
	L'~=~\left( \begin{tabular}{cccc}
	$\alpha_{2}$ & $0$ & $0$ & $0$\\
	$0$ & $\beta_{2}$ & $0$ & $0$ \\
	$0$ & $0$ & $\gamma_{2}$ & $0$ \\
	$0$ & $0$ & $1$ & $\gamma_{2}$\\
	\end{tabular} \right).
\end{eqnarray}
This solution has been denoted by $L^{(h_{4})}_{_{2}}$ in Table 1.
In this manner, we give the following theorem as one of the main results of this paper.
\\
\\
{\bf Theorem.} {\it Any algebraic Haantjes structure of the $h_{4}$ Lie algebra belongs just to one of the following 49 inequivalent classes of Table 1 \footnote{Note that using \eqref{2.5} one can obtain the Haantjes structures on the $H_{4}$ group manifold.}.}

\newpage
\begin{center}
\small {{{\bf Table 1.} The algebraic Haantjes structures on the $h_{4}$ Lie algebra}}
		{\scriptsize
			\renewcommand{\arraystretch}{1.5}{
\begin{tabular}{|p{0.8cm}|l|} \hline \hline
 Symbol & ~~Non-zero components of the inequivalent Haantjes structures  \\ \hline
{$L^{(h_{4})}_{_{1}}$} & ${L}^1_{~1}={L}^2_{~2}=\alpha_{1},~~~{L}^3_{~3}={L}^4_{~4}=\beta_{1},~~~{L}^2_{~1}={L}^3_{~1}={L}^4_{~2}={L}^4_{~3}=1$  \\\hline
	
{$L^{(h_{4})}_{_{2}}$} & ${L}^1_{~1}=\alpha_{2},~~~{L}^2_{~2}=\beta_{2}~~~{L}^3_{~3}={L}^4_{~4}=\gamma_{2},~~~{L}^4_{~3}=1$  \\\hline

{$L^{(h_{4})}_{_{3}}$} & ${L}^1_{~1}=\alpha_{3},~~~{L}^2_{~2}={L}^3_{~3}={L}^4_{~4}=\beta_{3},~~~{L}^4_{~2}={L}^4_{~3}=1$  \\\hline

{$L^{(h_{4})}_{_{4}}$} & ${L}^1_{~1}={L}^2_{~2}=\alpha_{4},~~~{L}^3_{~3}={L}^4_{~4}=\beta_{4},~~~{L}^2_{~3}={L}^3_{~1}={L}^4_{~2}=1$  \\\hline

{$L^{(h_{4})}_{_{5}}$} & ${L}^1_{~1}={L}^2_{~2}={L}^4_{~4}=\alpha_{5},~~~{L}^3_{~3}=\beta_{5},~~~{L}^2_{~3}={L}^4_{~2}=1$  \\\hline

{$L^{(h_{4})}_{_{6}}$} &${L}^3_{~3}=\alpha_{6},~~~{L}^1_{~1}={L}^4_{~4}=\beta_{6},~~~{L}^2_{~2}=-\alpha_{6}+2\beta_{6},~~~{L}^3_{~2}=-(\beta_{6}-\alpha_{6})^{2},~~~
{L}^2_{~3}=1$  \\\hline

{$L^{(h_{4})}_{_{7}}$} & ${L}^3_{~3}={L}^4_{~4}=\alpha_{7},~~~{L}^1_{~2}={L}^4_{~3}=1$  \\\hline

{$L^{(h_{4})}_{_{8}}$} & ${L}^3_{~3}={L}^4_{~4}=\alpha_{8},~~~{L}^2_{~1}=\alpha_{8}^{2}-\alpha_{8},~~~{L}^1_{~2}={L}^2_{~2}={L}^4_{~3}=1$  \\\hline

{$L^{(h_{4})}_{_{9}}$} & $\frac{1}{2}{L}^1_{~1}={L}^3_{~3}={L}^4_{~4}=\alpha_{9},~~~{L}^2_{~1}=-\alpha_{9}^{2},~~~{L}^1_{~2}={L}^4_{~2}={L}^4_{~3}=1$  \\\hline

{$L^{(h_{4})}_{_{10}}$} & ${L}^1_{~1}=\alpha_{_{10}}+\beta_{_{10}},~~~{L}^3_{~3}=\alpha_{_{10}},~~~{L}^4_{~4}=\beta_{_{10}},~~~{L}^2_{~1}=-\alpha_{_{10}}\beta_{_{10}},
~~~{L}^4_{~3}=\beta_{_{10}}( -\alpha_{_{10}}+\beta_{_{10}}),~~~{L}^1_{~2}={L}^4_{~2}=1$  \\\hline

{$L^{(h_{4})}_{_{11}}$} & ${L}^1_{~1}={L}^2_{~2}=\frac{1}{2}{L}^3_{~3}=\alpha_{_{11}},~~~{L}^4_{~3}=-\alpha_{_{11}}^{2},~~~{L}^2_{~1}={L}^3_{~4}=1$  \\\hline

{$L^{(h_{4})}_{_{12}}$} & ${L}^2_{~2}={L}^4_{~4}=\alpha_{_{12}},~~~{L}^1_{~3}={L}^4_{~2}=1$   \\\hline

{$L^{(h_{4})}_{_{13}}$} & ${L}^2_{~2}={L}^4_{~4}=\alpha_{_{13}},~~~{L}^3_{~1}=\alpha_{_{13}}^{2},~~~{L}^1_{~3}={L}^4_{~2}={L}^4_{~3}=1$  \\\hline

{$L^{(h_{4})}_{_{14}}$} & ${L}^2_{~2}={L}^4_{~4}=\alpha_{_{14}},~~~{L}^3_{~1}=\alpha_{_{14}}^{2},~~~{L}^1_{~3}={L}^4_{~2}=1$  \\\hline

{$L^{(h_{4})}_{_{15}}$} & $\frac{1}{2}{L}^1_{~1}={L}^2_{~2}={L}^4_{~4}=\alpha_{_{15}}~~~{L}^3_{~1}=-\alpha_{_{15}}^{2}~~~{L}^1_{~3}={L}^4_{~2}=1$  \\\hline

{$L^{(h_{4})}_{_{16}}$} & ${L}^1_{~1}={L}^2_{~2}=\alpha_{_{16}},~~~{L}^4_{~4}=\beta_{_{16}},~~~{L}^3_{~1}={L}^4_{~2}=\beta_{_{16}}(-\alpha_{_{16}}+\beta_{_{16}}),~~~{L}^1_{~3}={L}^2_{~3}=1$  \\\hline

{$L^{(h_{4})}_{_{17}}$} & $\frac{1}{2}{L}^1_{~1}={L}^2_{~2}={L}^4_{~4}=\alpha_{_{17}},~~~{L}^3_{~1}=-\alpha_{_{17}}^{2},~~~{L}^1_{~3}={L}^3_{~2}=1$  \\\hline

{$L^{(h_{4})}_{_{18}}$} & ${L}^1_{~1}={L}^3_{~3}=\alpha_{_{18}},~~~{L}^2_{~2}={L}^4_{~4}=\beta_{_{18}},~~~{L}^3_{~1}=(\alpha_{_{18}}-\beta_{_{18}})^{2},~~~
{L}^3_{~2}=-\alpha_{_{18}}+\beta_{_{18}},~~~{L}^1_{~2}={L}^1_{~3}=1$  \\\hline

{$L^{(h_{4})}_{_{19}}$} & ${L}^1_{~1}={L}^3_{~3}=\alpha_{_{19}},~~{L}^4_{~4}=\beta_{_{19}},~~
{L}^2_{~2}=2\alpha_{_{19}}-\beta_{_{19}},~~{L}^3_{~1}=(\alpha_{_{19}}-\beta_{_{19}})^{2},~~{L}^4_{~1}=4(\beta_{_{19}}-\alpha_{_{19}})^{3},~~{L}^1_{~2}={L}^1_{~3}=1$  \\\hline
{$L^{(h_{4})}_{_{20}}$} & ${L}^2_{~2}=\alpha_{_{20}},~~~\frac{1}{2}{L}^3_{~3}={L}^3_{~2}={L}^4_{~4}=\beta_{_{20}},~~~{L}^3_{~1}=-\beta_{_{20}}^{2},~~~{L}^1_{~2}={L}^1_{~3}=1$  \\\hline

{$L^{(h_{4})}_{_{21}}$} & ${L}^2_{~2}={L}^3_{~2}={L}^4_{~4}=\alpha_{_{21}},~~~{L}^3_{~1}=\alpha_{_{21}}^{2},~~~{L}^1_{~2}={L}^1_{~3}=1$  \\\hline

{$L^{(h_{4})}_{_{22}}$} & $-{L}^1_{~2}=-{L}^2_{~1}={L}^4_{~4}=\alpha_{_{22}},~~~-2{L}^3_{~1}={L}^3_{~2}=\frac{2}{3}{L}^4_{~1}=\frac{1}{3}{L}^4_{~2}=-\frac{1}{2}\alpha_{22}^{2},~~~{L}^1_{~3}={L}^2_{~3}=1$  \\\hline

{$L^{(h_{4})}_{_{23}}$} & $-{L}^2_{~1}={L}^4_{~4}=\alpha_{_{23}},~~~{L}^1_{~2}=1-\alpha_{_{23}},~~~2{L}^3_{~1}=-{L}^2_{~3}=\frac{1}{2}(-1+\alpha_{_{23}})^{2},$\\
 & $~{L}^4_{~1}=-\frac{3}{4}\alpha_{_{23}}^{2}-\frac{1}{2}\alpha_{_{23}}+\frac{1}{4},~~~
 {L}^4_{~2}=-\frac{3}{2}\alpha_{_{23}}^{2}+2\alpha_{_{23}}-\frac{1}{2},~~~{L}^1_{~3}={L}^2_{~2}={L}^2_{~3}=1$  \\\hline

{$L^{(h_{4})}_{_{24}}$} & ${L}^2_{~2}={L}^4_{~4}=\alpha_{_{24}},~~~{L}^3_{~1}=-\frac{1}{4}\beta_{_{24}}^{2},~~~{L}^1_{~1}=-2{L}^3_{~2}=\beta_{_{24}},~~~{L}^1_{~2}={L}^1_{~3}=1$  \\\hline
{$L^{(h_{4})}_{_{25}}$} & ${L}^1_{~1}=\beta_{_{25}},~~~{L}^2_{~2}={L}^4_{~4}=\alpha_{_{25}},~~~{L}^3_{~1}=-\frac{1}{4}(\beta_{_{25}}-1)^{2},~~~{L}^3_{~2}=-\frac{1}{2}\beta_{_{25}}+\frac{1}{2},~~~{L}^1_{~2}={L}^1_{~3}={L}^3_{~3}=1$  \\\hline
{$L^{(h_{4})}_{_{26}}$} & $\frac{1}{2}{L}^1_{~1}=-{L}^3_{~2}={L}^4_{~4}=\alpha_{_{26}},~~~{L}^3_{~1}=-\alpha_{_{26}}^{2},~~~{L}^1_{~2}={L}^1_{~3}=1$  \\\hline
{$L^{(h_{4})}_{_{27}}$} & ${L}^1_{~1}=\beta_{_{27}},~~{L}^4_{~4}=\alpha_{_{27}},~~{L}^1_{~2}=1-\alpha_{_{27}},~~{L}^2_{~1}=\beta_{_{27}}-\alpha_{_{27}},~~
{L}^3_{~1}=-(\beta_{_{27}}-2\alpha_{_{27}}+1)^{2},~~{L}^3_{~3}=-\beta_{_{27}}+3\alpha_{_{27}}-1,$\\
 &  $~{L}^1_{~3}={L}^2_{~2}={L}^2_{~3}=1$  \\\hline
 {$L^{(h_{4})}_{_{28}}$} & ${L}^1_{~1}=\beta_{_{28}},~~~{L}^4_{~4}=-{L}^1_{~2}=\alpha_{_{28}},~~~{L}^2_{~1}=\beta_{_{28}}-\alpha_{_{28}},~~~
 {L}^3_{~1}=-(\beta_{_{28}}-2\alpha_{_{28}})^{2},~~~{L}^3_{~3}=-\beta_{_{28}}+3\alpha_{_{28}}$,\\
  &   $~{L}^1_{~3}={L}^2_{~3}=1$  \\\hline
{$L^{(h_{4})}_{_{29}}$} & ${L}^1_{~1}=2\alpha_{_{29}}-\beta_{_{29}}-1,~~{L}^4_{~4}=\alpha_{_{29}},~~{L}^2_{~1}=-(\beta_{_{29}}-\alpha_{_{29}})^{2},
~~{L}^2_{~2}=\beta_{_{29}},~~{L}^3_{~1}=\alpha_{_{29}}-\beta_{_{29}}-1,~~{L}^3_{~3}=\alpha_{_{29}}+1,$\\
   &   ${L}^1_{~2}={L}^1_{~3}={L}^3_{~2}=1$  \\\hline
{$L^{(h_{4})}_{_{30}}$} & ${L}^1_{~1}=2\alpha_{_{30}}-\beta_{_{30}},~~~{L}^2_{~1}=-(\beta_{_{30}}-\alpha_{_{30}})^{2},~~~{L}^2_{~2}=\beta_{_{30}},
~~~{L}^3_{~3}={L}^4_{~4}=\alpha_{_{30}},~~~{L}^1_{~2}={L}^1_{~3}=1$  \\\hline
 \end{tabular}}}
\end{center}
$\\$
\\
\begin{center}
\small {{{\bf Table 1.}~ Continued}}
		{\scriptsize
			\renewcommand{\arraystretch}{1.5}{
\begin{tabular}{|p{0.8cm}|l|} \hline \hline
 Symbol& ~~Non-zero components of the inequivalent Haantjes structures  \\ \hline
 {$L^{(h_{4})}_{_{31}}$} & $\frac{1}{2}{L}^3_{~3}={L}^2_{~2}={L}^4_{~4}=\alpha_{_{31}},~~~{L}^3_{~1}=-\alpha_{_{31}}^{2},~~~{L}^1_{~2}={L}^1_{~3}=1$  \\\hline
{$L^{(h_{4})}_{_{32}}$} & ${L}^1_{~1}=2{L}^2_{~2}=2{L}^4_{~4}=-2{L}^3_{~2}=\alpha_{_{32}},~~~4{L}^3_{~1}=-\alpha_{_{32}}^{2},~~~{L}^1_{~2}={L}^1_{~3}=1$  \\\hline
{$L^{(h_{4})}_{_{33}}$} & ${L}^1_{~1}=\alpha_{_{33}},~~~{L}^2_{~2}={L}^4_{~4}=\frac{1}{2}\alpha_{_{33}}+\frac{1}{2},~~~{L}^3_{~1}=-\frac{1}{4}(\alpha_{_{33}}-1)^{2},
~~~{L}^3_{~2}=-\frac{1}{2}\alpha_{_{33}}+\frac{1}{2},~~~{L}^1_{~2}={L}^1_{~3}={L}^3_{~3}=1$  \\\hline
{$L^{(h_{4})}_{_{34}}$} & ${L}^1_{~1}=\alpha_{_{34}},~~~{L}^2_{~2}=\frac{1}{2}{L}^3_{~3}=\beta_{_{34}},~~~{L}^2_{~1}=\beta_{_{34}}(\beta_{_{34}}-\alpha_{_{34}}),
~~~{L}^4_{~3}=-\beta_{_{34}}^{2},~~~{L}^3_{~4}=1$  \\\hline
{$L^{(h_{4})}_{_{35}}$} & ${L}^1_{~1}={L}^2_{~2}={L}^4_{~4}=\alpha_{_{35}},~~~{L}^3_{~3}=\alpha_{_{35}}+1,~~~{L}^1_{~4}={L}^2_{~4}={L}^4_{~3}=1$  \\\hline
{$L^{(h_{4})}_{_{36}}$} & ${L}^1_{~1}={L}^2_{~2}={L}^4_{~4}=\beta_{_{36}},~~~{L}^3_{~3}=\alpha_{_{36}},~~~{L}^1_{~3}={L}^1_{~4}={L}^2_{~3}=1$  \\\hline
{$L^{(h_{4})}_{_{37}}$} & ${L}^1_{~1}={L}^3_{~3}=\alpha_{_{37}},~~~{L}^2_{~2}={L}^4_{~4}=\beta_{_{37}},~~~{L}^1_{~4}=\alpha_{_{37}}-\beta_{_{37}},~~~{L}^2_{~1}={L}^2_{~4}={L}^4_{~3}=1$  \\\hline
{$L^{(h_{4})}_{_{38}}$} & $\frac{1}{2}{L}^1_{~1}={L}^2_{~2}=\alpha_{_{38}},~~~{L}^3_{~3}=\beta_{_{38}},~~~{L}^4_{~1}=-\alpha_{_{38}}^{2},~~~{L}^1_{~4}={L}^4_{~2}=1$  \\\hline
{$L^{(h_{4})}_{_{39}}$} & ${L}^3_{~3}=\alpha_{_{39}},~~~{L}^1_{~4}=-\frac{1}{4},~~~{L}^2_{~2}=\frac{1}{2},~~~{L}^4_{~1}={L}^4_{~2}=1$  \\\hline
{$L^{(h_{4})}_{_{40}}$} & ${L}^3_{~3}=\alpha_{_{40}},~~~{L}^1_{~2}={L}^1_{~3}={L}^1_{~4}=1$  \\\hline
{$L^{(h_{4})}_{_{41}}$} & ${L}^1_{~2}={L}^2_{~1}=-{L}^4_{~4}=-\alpha_{_{41}},~~~\frac{1}{4}{L}^3_{~1}=-\frac{1}{2}{L}^3_{~2}=\frac{1}{3}{L}^4_{~1}=-\frac{1}{3}{L}^4_{~2}=\alpha_{41}^{2},~~~{L}^1_{~3}={L}^2_{~3}=1$  \\\hline
{$L^{(h_{4})}_{_{42}}$} & ${L}^1_{~1}=\alpha_{_{42}},~~~{L}^2_{~2}={L}^3_{~3}={L}^4_{~4}=\beta_{_{42}},~~~{L}^2_{~3}=1$  \\\hline
{$L^{(h_{4})}_{_{43}}$} & ${L}^1_{~2}={L}^1_{~3}={L}^4_{~3}=1 $ \\\hline

{$L^{(h_{4})}_{_{44}}$} &  ${L}^1_{~2}={L}^1_{~3}={L}^3_{~4}=1$   \\\hline
	
{$L^{(h_{4})}_{_{45}}$} & ${L}^1_{~2}={L}^1_{~3}={L}^1_{~4}=1$  \\\hline

{$L^{(h_{4})}_{_{46}}$} & ${L}^1_{~1}={L}^1_{~2}={L}^1_{~3}={L}^1_{~4}={L}^2_{~2}={L}^3_{~3}={L}^4_{~4}=1$  \\\hline
$L^{(h_{4})}_{_{47}}$&${L}^1_{~2}=-{L}^3_{~4}=\alpha_{_{47}}$    \\\hline

$L^{(h_{4})}_{_{48}}$&$-{L}^1_{~2}=-{L}^1_{~3}={L}^2_{~2}={L}^2_{~4}={L}^3_{~1}={L}^3_{~3}=-{L}^4_{~2}=-{L}^4_{~4}=\alpha_{_{48}},~~~
\frac{1}{2}{L}^2_{~1}={L}^2_{~3}=-{L}^4_{~1}=\beta_{_{48}}$   \\\hline
	
$L^{(h_{4})}_{_{49}}$&$-{L}^1_{~1}={L}^4_{~4}=\frac{\lambda \gamma_{_{49}}}{2},~~~{L}^1_{~4}=\gamma_{_{49}},~~~{L}^4_{~1}=-\frac{\lambda^{2}\gamma_{_{49}}}{4},~{L}^3_{~1}={L}^4_{~2}=\alpha_{_{49}},~~~
{L}^3_{~2}=\beta_{_{49}},~~~{L}^1_{~2}=-{L}^3_{~4}=\frac{2\alpha_{_{49}}}{\lambda}$    \\\hline
 \end{tabular}}}
\end{center}
\vspace{-3mm}
$~~~~~~~~~${\scriptsize where $\alpha_{i}, \beta_{i}, \gamma_{i}$ and $\lambda$  are some real constants.}\\
Below we will discuss under what conditions the resulting Haantjes structures are equivalent. Using the transformation \eqref{3.2} and formula \eqref{2.16}
we find that:
\begin{itemize}
\item The Haantjes structures $L^{(h_{4})}_{_{6}}$ and $L^{(h_{4})}_{_{42}}$ are equivalent to each other if one sets $\alpha_{6}=\alpha_{_{42}}=\beta_{6}=\beta_{_{42}}$. Otherwise, they are inequivalent.

\item The Haantjes structures $L^{(h_{4})}_{_{10}}$, $L^{(h_{4})}_{_{22}}$, $L^{(h_{4})}_{_{28}}$ and $L^{(h_{4})}_{_{41}}$ are equivalent to each other if
one sets $\alpha_{_{10}}=\alpha_{_{22}}=\alpha_{_{28}}=\alpha_{_{41}}$,  $\beta_{_{10}}=0$ and $\beta_{_{28}}=d$.

\item Structures $L^{(h_{4})}_{_{7}}$, $L^{(h_{4})}_{_{9}}$, $L^{(h_{4})}_{_{13}}$ and $L^{(h_{4})}_{_{14}}$ are equivalent to each other if one sets
    $\alpha_{_{7}}=\alpha_{_{9}}=\alpha_{_{13}}=\alpha_{_{14}}=0$.

\item Structures $L^{(h_{4})}_{_{18}}$, $L^{(h_{4})}_{_{19}}$, $L^{(h_{4})}_{_{20}}$, $L^{(h_{4})}_{_{24}}$, $L^{(h_{4})}_{_{25}}$, $L^{(h_{4})}_{_{27}}$,
$L^{(h_{4})}_{_{28}}$, $L^{(h_{4})}_{_{29}}$, $L^{(h_{4})}_{_{30}}$, $L^{(h_{4})}_{_{32}}$, $L^{(h_{4})}_{_{33}}$ and $L^{(h_{4})}_{_{35}}$
are equivalent to each other if one sets $\alpha_{18}=\alpha_{19}=\alpha_{20}=\alpha_{24}=\alpha_{25}=\alpha_{27}=\alpha_{28}=\alpha_{29}=\alpha_{30}$,
$\beta_{18}=\beta_{19}=\beta_{20}=\beta_{29}=\beta_{30}$, $\beta_{24}=\beta_{28}=\beta_{32}=2\beta_{18}$, $\beta_{25}=\alpha_{35}=-2d+1$ and $\beta_{27}=\alpha_{33}=-1$.

\item Structures $L^{(h_{4})}_{_{22}}$, $L^{(h_{4})}_{_{28}}$ and $L^{(h_{4})}_{_{41}}$ are equivalent to each other if one sets $\alpha_{22}=\alpha_{28}=\alpha_{41}=0$.

\item Structures $L^{(h_{4})}_{_{20}}$, $L^{(h_{4})}_{_{24}}$, $L^{(h_{4})}_{_{27}}$, $L^{(h_{4})}_{_{28}}$, $L^{(h_{4})}_{_{30}}$ and $L^{(h_{4})}_{_{32}}$
are equivalent to each other if one sets
    $\alpha_{20}=\alpha_{24}=\alpha_{27}=\alpha_{28}=\alpha_{30}=\beta_{20}=\beta_{30}$,
    $\beta_{28}=\alpha_{32}=2\beta_{20}$, $\beta_{24}=2\beta_{20}-1$ and $\beta_{27}=\beta_{20}-\frac{1}{a}$.

\item Structures $L^{(h_{4})}_{_{24}}$, $L^{(h_{4})}_{_{27}}$, $L^{(h_{4})}_{_{28}}$, $L^{(h_{4})}_{_{29}}$ and $L^{(h_{4})}_{_{30}}$
are equivalent to each other if one sets
    $\alpha_{24}=\alpha_{27}=\alpha_{28}=\alpha_{29}=\alpha_{30}=\beta_{29}=\beta_{30}$,
    $\beta_{24}=\beta_{28}=2\alpha_{28}$ and $\beta_{27}=\frac{a-2}{a}$.

\item Structures $L^{(h_{4})}_{_{28}}$, $L^{(h_{4})}_{_{29}}$, $L^{(h_{4})}_{_{30}}$ and $L^{(h_{4})}_{_{33}}$ are equivalent to each other if one sets
    $\alpha_{28}=\alpha_{29}=\alpha_{30}=\beta_{29}=\beta_{30}=c$,
   $\beta_{28}=2\alpha_{28}$ and $\alpha_{33}=\beta_{28}-1$.

\item Structures $L^{(h_{4})}_{_{21}}$, $L^{(h_{4})}_{_{24}}$, $L^{(h_{4})}_{_{25}}$, $L^{(h_{4})}_{_{26}}$, $L^{(h_{4})}_{_{27}}$, $L^{(h_{4})}_{_{29}}$,
$L^{(h_{4})}_{_{30}}$, $L^{(h_{4})}_{_{32}}$ and $L^{(h_{4})}_{_{33}}$ are equivalent to each other if one sets
    $\alpha_{21}=\alpha_{24}=\alpha_{25}=\alpha_{26}=\alpha_{27}=\alpha_{29}=\alpha_{30}=\alpha_{32}=0$,
     $\beta_{24}=\beta_{29}=\beta_{30}=0$ and $\beta_{25}=\beta_{27}=\alpha_{33}=-1$.

\item Structures $L^{(h_{4})}_{_{26}}$, $L^{(h_{4})}_{_{27}}$, $L^{(h_{4})}_{_{29}}$, $L^{(h_{4})}_{_{30}}$, $L^{(h_{4})}_{_{32}}$ and
$L^{(h_{4})}_{_{33}}$ are equivalent to each other if one sets
    $\alpha_{26}=\alpha_{27}=\alpha_{29}=\alpha_{30}=\alpha_{32}=0$,
     $\beta_{29}=\beta_{30}=0$ and $\beta_{27}=\alpha_{33}=-1$.

\item Structures $L^{(h_{4})}_{_{25}}$, $L^{(h_{4})}_{_{28}}$, $L^{(h_{4})}_{_{29}}$ and $L^{(h_{4})}_{_{33}}$
are equivalent to each other if one sets $\alpha_{25}=\alpha_{28}=\alpha_{29}=\beta_{29}$, $\beta_{25}=\alpha_{33}=\beta_{28}-1$ and $\beta_{28}=2\alpha_{28}$.

\item Structures $L^{(h_{4})}_{_{10}}$ and $L^{(h_{4})}_{_{15}}$ are equivalent to each other if one sets $\alpha_{10}=\alpha_{15}=\beta_{10}$.

\item Structures $L^{(h_{4})}_{_{10}}$ and $L^{(h_{4})}_{_{16}}$ are equivalent to each other if one sets $\alpha_{10}=\alpha_{16}$ and $\beta_{10}=\beta_{16}=0$.

\item Structures $L^{(h_{4})}_{_{10}}$ and $L^{(h_{4})}_{_{27}}$ are equivalent to each other if one sets $\alpha_{10}=\alpha_{27}=\beta_{10}=1$ and $\beta_{27}=d+2$.

\item Structures $L^{(h_{4})}_{_{18}}$ and $L^{(h_{4})}_{_{41}}$ are equivalent to each other if one sets $\alpha_{18}=\frac{1}{2a}$ and $\alpha_{41}=\beta_{18}=-\frac{1}{a}$.

\item Structures $L^{(h_{4})}_{_{20}}$ and $L^{(h_{4})}_{_{23}}$ are equivalent to each other if one sets $\alpha_{20}=\alpha_{23}=\beta_{20}=\frac{1}{3}$.

\item Structures $L^{(h_{4})}_{_{22}}$ and $L^{(h_{4})}_{_{24}}$ are equivalent to each other if one sets $\alpha_{22}=\alpha_{24}=-\beta_{24}=c$.

\item Structures $L^{(h_{4})}_{_{22}}$ and $L^{(h_{4})}_{_{25}}$ are equivalent to each other if one sets
$\alpha_{22}=\alpha_{25}=c$ and $\beta_{25}=-c-1$.

\item Structures $L^{(h_{4})}_{_{23}}$ and $L^{(h_{4})}_{_{24}}$ are equivalent to each other if one sets $\alpha_{23}=\alpha_{24}=c+1$
and $\beta_{24}=-c$.

\item Structures $L^{(h_{4})}_{_{23}}$ and $L^{(h_{4})}_{_{25}}$ are equivalent to each other if one sets $\alpha_{23}=\alpha_{25}=-\beta_{25}=1+c$.

\item Structures $L^{(h_{4})}_{_{23}}$ and $L^{(h_{4})}_{_{27}}$ are equivalent to each other if one sets $\alpha_{23}=\alpha_{27}=-\beta_{27}=\frac{1}{3}$.

\item Structures $L^{(h_{4})}_{_{23}}$ and $L^{(h_{4})}_{_{28}}$ are equivalent to each other if one sets $\alpha_{23}=\alpha_{28}=\frac{1}{3}$ and $\beta_{28}=\frac{2}{3}$.

\item Structures $L^{(h_{4})}_{_{23}}$ and $L^{(h_{4})}_{_{29}}$ are equivalent to each other if one sets $\alpha_{23}=\alpha_{29}=\beta_{29}=\frac{1}{3}$.

\item Structures $L^{(h_{4})}_{_{23}}$ and $L^{(h_{4})}_{_{33}}$ are equivalent to each other if one sets $\alpha_{23}=-\alpha_{33}=\frac{1}{3}$.

\item Structures $L^{(h_{4})}_{_{27}}$ and $L^{(h_{4})}_{_{32}}$ are equivalent to each other if one sets $\alpha_{32}=2\alpha_{27}=-2d$
and $\beta_{27}=-2d-1$.
\end{itemize}
\section{Integrable sigma model with Haantjes structure on a Lie group }

\subsection {A review of the integrability of two-dimensional sigma model }

In this subsection, let us review the notations and key aspects of the general formalism introduced by Mohammedi \cite{Exact} for ensuring the integrability of a sigma model on a manifold.
To illustrate these concepts, let us consider the following two-dimensional sigma model action
\begin{equation}\label{eqn:sm}
S=\int_{\Sigma}\hspace{2mm}dzd\overline{z}\big(G_{\mu\nu}(x)+B_{\mu\nu}(x)\big)\hspace{0mm}\partial\hspace{0mm}x^{\mu}\overline{\partial}x^{\nu},
\end{equation}
where $x^{\mu}(z,\overline{z}), \mu=1,2,...,m$ are the coordinates of $m$-dimensional manifold $M$, with $G_{\mu\nu}$ and $B_{\mu\nu}$ as metric and B-field on it. Here, $(z,\overline{z})$ are the complex coordinates which are defined by $(z=\tau +i \sigma, \bar{z}=\tau -i \sigma)$ together with
$\partial = \frac{\partial}{\partial z}$ and ${\bar \partial} = \frac{\partial}{\partial \bar{z}}$, where $(\tau , \sigma)$ are
the two-dimensional coordinates of the world-sheet $\Sigma$. The equations of motion of this model have the following form \cite{Exact}
\begin{equation}\label{eom}
  \overline{\partial} \partial\hspace{0cm}x^{\lambda}+\Omega^{\lambda}\hspace{0cm}_{\mu\nu}\partial\hspace{0cm}x^{\mu}\overline{\partial}x^{\nu}=0,
\end{equation}
where
\begin{equation}
 \Omega^{\lambda}\hspace{0cm}_{\mu\nu}=\Gamma^{\lambda}\hspace{0cm}_{\mu\nu}-H^{\lambda}\hspace{0cm}_{\mu\nu},
\end{equation}
in which $\Gamma^{\lambda}\hspace{0cm}_{\mu\nu}$ are
Christoffel coefficients and the components of torsion $H^{\lambda}\hspace{0cm}_{\mu\nu}$ defined by

\begin{equation}
H^{\lambda}\hspace{0cm}_{\mu\nu}=\frac{1}{2}G^{\lambda\eta}(\partial_{\eta}B_{\mu\nu}+\partial_{\nu}B_{\eta\mu}+\partial_{\mu}B_{\nu\eta}).
\end{equation}
According to \cite{Exact}, one can construct the following linear system, whose consistency
conditions (a zero curvature representation) are equivalence to
the equations of motion (\ref{eom})
\begin{align}
[\partial+\partial\hspace{0cm}x^{\mu} \alpha_{\mu}(x , \lambda)]\psi=0,
\notag
\end{align}
\begin{equation}\label{lax}
[\overline{\partial}+\overline{\partial}x^{\mu} \beta_{\mu}(x , \lambda)]\psi=0,
\end{equation}
where $\alpha_{\mu}(x , \lambda)$ and $\beta_{\mu}(x , \lambda)$ are matrix functions of coordinates $x^{\mu}$, and $\lambda$ is a spectral parameter
which plays an important role in the construction of conserved quantities of integrable sigma models.
The compatibility condition of this linear system yields the
equations of motion if the matrices $\alpha_{\mu}(x)$
and $\beta_{\mu}(x)$ satisfies the following relations \cite{Exact}
\begin{equation}\label{me1}
 \partial\hspace{0cm}_{\mu}\beta_{\nu}-\partial\hspace{0cm}_{\nu}\alpha_{\mu}+[\alpha_{\mu},\beta_{\nu}]=\Omega^{\lambda}\hspace{0cm}_{\mu\nu}\mu_{\lambda},
\end{equation}
where $\mu_{\mu}=\beta_{\mu}-\alpha_{\mu}$. One may write the equation (\ref{me1}) as
\begin{equation}\label{me}
 F_{\mu\nu}=-(\nabla_{\mu}\mu_{\nu}-\Omega^{\lambda}\hspace{0cm}_{\mu\nu}\mu_{\lambda}),
\end{equation}
where the field strength $F_{\mu\nu}$ and covariant derivative with respect to the matrices $\alpha_{\mu}$ are given by
\begin{equation}
 F_{\mu\nu}=\partial_{\mu}\alpha_{\nu}-\partial_{\nu}\alpha_{\mu}+[\alpha_{\mu},\alpha_{\nu}],~~~~~~~~~~~~~~~~~~
 \nabla_{\mu} {\cal O}=\partial_{\mu}{\cal O}+[\alpha_{\mu},{\cal O}].
\end{equation}
Note that by splitting the equation (\ref{me}) into its symmetric and anti-symmetric components, we can express it as the following set of relations{\footnote{Note that the equation (\ref{eqn:sp}) is a gauged version of a matrix-valued Killing equation. Indeed, if $[\alpha_{\mu},\mu_{\nu}]+[\alpha_{\nu},\mu_{\mu}]=0$, then this equation is simplified as Killing one,  $\partial_{\mu}\mu_{\nu}+\partial_{\nu}\mu_{\mu}-2\Gamma^{\lambda}_{\mu\nu}\mu_{\lambda}=0$ \cite{Exact}.}}
\begin{eqnarray}
 0&=&\nabla_{\mu}\mu_{\nu}+\nabla_{\nu}\mu_{\mu}-2\Gamma^{\lambda}\hspace{0cm}_{\mu\nu}\mu_{\lambda},\label{eqn:sp}\\
 F_{\mu\nu}&=&-\frac{1}{2}(\nabla_{\mu}\mu_{\nu}-\nabla_{\nu}\mu_{\mu})-H^{\lambda}\hspace{0cm}_{\mu\nu}\mu_{\lambda}.\label{eqn:asp}
\end{eqnarray}
Thus, the integrability condition of the sigma model (\ref{eqn:sm}) is equivalent to finding matrices $\alpha_{\mu}$ and $\mu_{\mu}$ that 
satisfy either the single relation (\ref{me}) or the pair of relations (\ref{eqn:sp}) and (\ref{eqn:asp}).
Note that these equations do not guarantee that the matrices $\alpha_{\mu}$ and $\mu_{\mu}$ depend on a spectral parameter. 
This dependency must be examined separately for the model.
We will apply the above formulations to examine the integrability conditions of a sigma model which will be introduced in the next subsection.

 \subsection{A new integrable sigma model with Haantjes structure}

 Now using the algebraic Haantjes structure we propose the following two-dimensional sigma model on a Lie group $G$:
\begin{eqnarray}
S=\int d^{2}z(g^{-1}\partial g)^{i}\big(\Omega_{ij}+a_{1}{ L}^k_{~i}\Omega_{kj}+ a_{2}{L}^k_{~i}{ L}^l_{~k}\Omega_{lj} +a_{3}{ L}^k_{~i}{ L}^l_{~k}{ L}^m_{~l}\Omega_{mj}\big)(g^{-1}\bar{\partial} g)^{j},\label{4.1}
\end{eqnarray}
where $g\in G$ and $a_{1}, a_{2}$ and $a_3$ being real constants; moreover, $\Omega_{ij}$ (being its inverse $\Omega^{ij}$)
stands for the ad-invariant metric on $\G$ of $G$. When $a_{1}=a_2=a_{3}=0$ we arrive at the chiral sigma model \cite{HE1}. 
In this manner we have deformed the chiral sigma model as in the model \eqref{4.1}. 
In the following we shall examine conditions on $\boldsymbol L$ in such a way that (while the $\boldsymbol L$ is Haantjes structure), the resulting model is integrable.
For this purpose, we consider the symmetric algebraic part of the above action as follows:
\begin{eqnarray}
&&G_{ij}=\Omega_{ij}+
\frac{1}{2}a_{1}({ L}^k_{~i}\Omega_{kj}+{L}^k_{~j}\Omega_{ki})+
\frac{1}{2}a_{2}({L}^k_{~i}{ L}^l_{~k}\Omega_{lj}+
 {L}^k_{~j}{ L}^l_{~k}\Omega_{li})\nonumber\\
 &&~~~~~~~~~~~~~~~~~~~~~~~~~~~~~~~~~~+\frac{1}{2}a_{3}({L}^k_{~i}{ L}^l_{~k}{ L}^m_{~l}\Omega_{mj}+ {L}^k_{~j}{ L}^l_{~k}{ L}^m_{~l}\Omega_{mi}),\label{4.2}
\end{eqnarray}
with the inverse proposed
\begin{eqnarray} \label{4.3}
&&G^{ij}=\Omega^{ij}+
\frac{1}{2}b_{1}({ L}^i_{~k}\Omega^{kj}+{L}^j_{~k} \Omega^{ki})+
\frac{1}{2}b_{2}({ L}^i_{~k} { L}^k_{~l}\Omega^{lj}+{ L}^j_{~k}{ L}^k_{~l}\Omega^{li})\nonumber\\
&&~~~~~~~~~~~~~~~~~~~~~~~~~~~~~~~~~+\frac{1}{2}b_{3}({ L}^i_{~k} { L}^k_{~l} { L}^l_{~m}\Omega^{mj}+{ L}^j_{~k}{ L}^k_{~l}{ L}^l_{~m}\Omega^{mi}),
\end{eqnarray}
while the antisymmetric part of the action or algebraic $B$-field can be written in the following form
\begin{eqnarray} \label{4.4}
&&B_{ij}=\frac{1}{2}a_{1}({ L}^k_{~i}\Omega_{kj}-{ L}^k_{~j}\Omega_{ki})+
\frac{1}{2}a_{2}({ L}^k_{~i}{ L}^l_{~k}\Omega_{lj}-{ L}^k_{~j}{ L}^l_{~k}\Omega_{li})\nonumber \\
&&~~~~~~~~~~~~~~~~~~~~~~~~~~~~~~~~~~~~~+\frac{1}{2}a_{3}({ L}^k_{~i}{ L}^l_{~k}{ L}^m_{~l}\Omega_{mj}-{ L}^k_{~j}{ L}^l_{~k}{ L}^m_{~l}\Omega_{mi}).
\end{eqnarray}
Using the invertibility of $G_{ij}$, we arrive at the following conditions
\begin{eqnarray}
a_{1}=-b_{1},a_{2}=b_{2}=0,a_{3}=b_{3}&=&0,\label{4.5}\\
\Omega L \Omega^{-1} L^{t}+L^{t}\Omega L \Omega^{-1} &=&0,\label{4.6}\\
(L^{t})^{2}+\Omega L^{2} \Omega^{-1}&=&0.\label{4.7}
\end{eqnarray}
Then, under the condition \eqref{4.5} the sigma model \eqref{4.1} becomes
\begin{eqnarray}
S=\int d^{2}z(g^{-1}\partial g)^{i}\big(\Omega_{ij}+a_{1}{ L}^k_{~i}\Omega_{kj}\big)(g^{-1}\bar{\partial} g)^{j},\label{4.7.1}
\end{eqnarray}
for which we find that
\begin{eqnarray}
G_{ij}&=&\Omega_{ij}+\frac{1}{2}a_{1}({ L}^k_{~i}\Omega_{kj}+{L}^k_{~j}\Omega_{ki}),\label{4.8}\\
B_{ij}&=&\frac{1}{2}a_{1}({ L}^k_{~i}\Omega_{kj}-{ L}^k_{~j}\Omega_{ki}).\label{4.9}
\end{eqnarray}
We see that the models with different ${ L}^i_{~j}$ have different metric and B-field.


 \subsubsection {The conditions for integrability of the sigma model }

In order to investigate the conditions under which the sigma model \eqref{4.7.1} is integrable we propose the $\alpha_{\mu}$ and $\beta_{\mu}$ matrices as follow:
\begin{eqnarray}
\alpha_{\mu}&=&e_{\mu}^{~~i} (c_{1}{ L}^j_{~i}+c_{2}{ L}^k_{~i}{ L}^j_{~k}+c_{3}{ L}^k_{~i}{ L}^l_{~k}{ L}^j_{~l})T_{j},\label{4.20}\\
\beta_{\mu}&=&e_{\mu}^{~~i} (d_{1}{ L}^j_{~i}+d_{2}{ L}^k_{~i}{ L}^j_{~k}+d_{3}{ L}^k_{~i}{ L}^l_{~k}{ L}^j_{~l})T_{j},\label{4.21}
\end{eqnarray}
where $c_i$ and $d_i$ are some real constants.
Here we define  $G_{\mu\nu}=e_{\mu}^{~~i} e_{\nu}^{~~j} G_{ij}$ and $B_{\mu\nu}=e_{\mu}^{~~i} e_{\nu}^{~~j} B_{ij}$ where $G_{ij}$ and $B_{ij}$ are given by \eqref{4.8} and \eqref{4.9}.
Then, imposing the conditions \eqref{eqn:sp} and \eqref{eqn:asp} and using the algebraic Haantjes relation, \eqref{2.11}, one
guarantees the integrability of the sigma model \eqref{4.7.1} provided that the following relations in the matrix form hold
\begin{eqnarray}
&& L \Omega^{-1} {\cal X}_{_k} L^{t}\Omega+L ({\cal X}_{_k})^{t} L+ L \Omega^{-1} {\cal Y}^{i} (\Omega L)_{ki}+L ({\cal X}_{_j})^{t} L^{j}_{~k}\nonumber\\
&&~~~~~~~~~~~~~~~~~~+L \Omega^{-1}L^{t}\Omega ({\cal X}_{_k})^{t}-L^{2} ({\cal X}_{_k})^{t}+
\frac{2c_{1}}{a_{1}}({\cal X}_{_j})^{t}L^{j}_{~k} L=0,\label{4.22}\\
&&a_{1}c_{1}\Big[-L^{2}\Omega^{-1} {\cal X}_{_k} L^{t}\Omega-L^{2}({\cal X}_{_k})^{t} L-L \Omega^{-1} L^{t} {\cal X}_{_k}L^{t}\Omega+L \Omega^{-1} L^{t}{\cal X}_{_k}\Omega L+L^{3} ({\cal X}_{_k})^{t}\nonumber\\
&&~~~~~~~~~~~~~~~~~~-2L^{2}\Omega^{-1} L^{t} \Omega ({\cal X}_{_k})^{t}+L \Omega^{-1} L^{t} \Omega L ({\cal X}_{_k})^{t}-L^{2}\Omega^{-1} {\cal Y}^{i} (\Omega L)_{ki}\nonumber\\
&&~~~~~~~~~~~~~~~~~~-L \Omega^{-1} L^{t}{\cal Y}^{i} (\Omega L)_{ki}+L \Omega L^{t} {\cal Y}^{i} (\Omega L)_{ik}-L^{2} ({\cal X}_{_j})^{t}L^{j}_{~k}\Big]
+2c_{2}\Big[L \Omega^{-1}{\cal X}_{_k} L^{t}\Omega\nonumber\\
&&~~~~~~~~~~~~~~~~~~+L^{2} ({\cal X}_{_k})^{t} L-L^{3} ({\cal X}_{_k})^{t}+L^{2}\Omega^{-1} {\cal Y}^{i} (\Omega L)_{ki}+L^{2} ({\cal X}_{_j})^{t} L^{j}_{~k}+L^{2} \Omega^{-1} L^{t}\Omega  ({\cal X}_{_k})^{t}\Big]\nonumber\\
&&~~~~~~~~~~~~~~~~~~+\frac{4c_{1}c_{2}}{a_{1}}\big[({\cal X}_{_j})^{t} L^{j}_{~k} L^{2}+({\cal X}_{_j})^{t} (L^{2})^{j}_{~k} L\big]=0,\label{4.23}\\
&&-L^{3} \Omega^{-1} {\cal X}_{_k} L^{t}\Omega-L^{3} ({\cal X}_{_k})^{t} L-L^{2} \Omega^{-1} L^{t} {\cal X}_{_k}L^{t}\Omega
\nonumber\\
&&~~~~~~~~~~~~~~~~~~~~+L^{2}\Omega^{-1} L^{t} {\cal X}_{_k}\Omega L+L^{4} {\cal X}_{_k}+L^{2}\Omega^{-1} L^{t} {\cal Y}^{i} (\Omega L)_{ik}\nonumber\\
&&~~~~~~~~~~~~~~~~~~~~+ L^{2}\Omega^{-1} L^{t} \Omega L ({\cal X}_{_k})^{t}-L^{3} {\cal Y}^{i} (\Omega L)_{ki}-L^{3} ({\cal X}_{_j})^{t}L^{j}_{~k}-L^{2}\Omega^{-1} L^{t} {\cal Y}^{i} (\Omega L)_{ki}\nonumber\\
&&~~~~~~~~~~~~~~~~~~~~-L^{3}\Omega^{-1} L^{t} \Omega ({\cal X}_{_k})^{t}
-L^{2}\Omega^{-1} (L^{t})^{2}\Omega ({\cal X}_{_k})^{t}+\frac{4c_{2}}{a_{1}^{2}} ({\cal X}_{_j})^{t} (L^{2})^{j}_{~k} L^{2}=0,\label{4.24}
\end{eqnarray}
where $({\cal Y}^{i})_{jk}=-{f^i}_{jk}$ are the matrix representations of the structure constants of $\G$.
In addition to the above relations,  the conditions $c_{i}=-d_{i}, (i=1, 2)$  and $c_3=d_3=0$ must be held between the coefficients 
$c_{i}$ and $d_{i}$\footnote{Considering $c_{i}=-d_{i}, (i=1, 2)$  and $c_3=d_3=0$, the Lax pairs of equations \eqref{4.20} and \eqref{4.21} are related to each other as
$\alpha_{\mu}=-\beta_{\mu}$, and hence that $\mu_{\mu} = \beta_{\mu} -\alpha_{\mu}$
defined below \eqref{me1} should become $\mu_{\mu} = -2\alpha_{\mu}$. So, according to the comment made in footnote 7,  
we arrive at the Killing equation for $\alpha_{\mu}$
so that the number of Killing vectors (dimension of isometry group) 
is equal to the background dimension of the group manifold $G$. This means that $G$ is an isometry group.}.

Note that here for simplicity of calculations we use the multiplicative spectral parameter in \eqref{4.20} and \eqref{4.21} as \cite{Exact}. 
Furthermore, in order to derive relations \eqref{4.22}-\eqref{4.24}  
we have compared the powers of $L$ appearing in relations \eqref{eqn:sp} and \eqref{eqn:asp}, meaning that first power of $L$ is equal to first on both sides of the equations, 
second powers of $L$ are equal to second of $L$, and so on.
In this manner, we found that the conditions $c_{i}=-d_{i}, (i=1, 2)$  and $c_3=d_3=0$ must be held between the coefficients $c_{i}$ and $d_{i}$.
Here, one can apply $c_1$ or $c_2$ as a spectral parameter for the model.
Thus, in general the Lax pairs of the model \eqref{4.7.1} become 
\begin{eqnarray}
\alpha_{\mu}&=&e_{\mu}^{~i} (c_{1}{ L}^j_{~i}+c_{2}{ L}^k_{~i}{ L}^j_{~k}) T_j,\label{4.26}\\
\beta_{\mu}&=&e_{\mu}^{~i} (d_{1}{ L}^j_{~i}+d_{2}{ L}^k_{~i}{ L}^j_{~k}) T_j.\label{4.27}
\end{eqnarray}
For a special case that ${ L}^j_{~i} = { \delta}^j_{~i}$, the model \eqref{4.7.1} turns into the principal chiral model \cite{KP, VG, HE1, HE2} with the Lax pairs 
$\alpha_{\mu}= (c_1 +c_2) e_{\mu}^{~i}  ~T_i$ and $\beta_{\mu}= (d_1 +d_2) e_{\mu}^{~i}~  T_i$
so that one can consider $c_1+c_2 = \frac{1}{1+\lambda}$ and $d_1+d_2 = \frac{1}{1-\lambda}$.
This shows that the Haantjes-deformed Lax connections are related to the ones of the undeformed principal chiral model.

\section{Integrable sigma models with the Haantjes structure on the $H_{4}$ Lie group }

In order to obtain the integrable sigma model \eqref{4.7.1} with the conditions \eqref{4.22}-\eqref{4.24} and also equations \eqref{4.5}-\eqref{4.7}
on the $H_{4}$ we must solve these relations together with \eqref{2.11}.
After solving the mentioned relations and imposing the conditions $c_{i}=-d_{i}, (i=1, 2)$ and  $c_{3}=d_{3}=0$, and then applying the Haantjes structures of Table 1
we find the algebraic Haantjes structures satisfying the integrability conditions.
The results show that among the Haantjes structures of Table 1, only three structures $L^{(h_{4})}_{_{47}}, L^{(h_{4})}_{_{48}}$
and $L^{(h_{4})}_{_{49}}$ satisfy the integrability conditions provided that $c_1$ is set zero. We have summarized them in Table 2.
For the structure $L^{(h_{4})}_{_{47}}$ we have set $\alpha_{_{47}} = \epsilon$, and for the $L^{(h_{4})}_{_{48}}$, $\alpha_{_{48}} =\delta$ and $\beta_{_{48}} =\xi$.
In order to satisfy the integrability conditions with the structure $L^{(h_{4})}_{_{49}}$, the $\lambda$ parameter must be equal to
the $\rho$ parameter appearing in the metric $\Omega_{ij}$ in \eqref{5.2}. In addition, we have set $\alpha_{_{49}} =\alpha$,  $\beta_{_{49}} =\beta$ and $\gamma_{_{49}} =\eta$.
Note that $\epsilon, \xi, \alpha, \beta, \delta, \eta $ and $\rho$ are some real constants.

\begin{center}
\small {{{\bf Table 2.} {\small Only the algebraic Haantjes structures on the $h_{4}$ satisfying the integrability conditions}}}\\
{\scriptsize \renewcommand{\arraystretch}{1.5}{
\begin{tabular}{|p{1cm}|l|} \hline \hline
 Symbol  & Non-zero components of the inequivalent Haantjes structures  \\ \hline

$L^{(h_{4})}_{_{47}}$&${L}^1_{~2}=-{L}^3_{~4}=\epsilon$    \\\hline

$L^{(h_{4})}_{_{48}}$&$-{L}^1_{~2}=-{L}^1_{~3}={L}^2_{~2}={L}^2_{~4}={L}^3_{~1}={L}^3_{~3}=-{L}^4_{~2}=-{L}^4_{~4}=\delta,~~~~
\frac{1}{2}{L}^2_{~1}={L}^2_{~3}=-{L}^4_{~1}=\xi$   \\\hline
	
$L^{(h_{4})}_{_{49}}$&$-{L}^1_{~1}={L}^4_{~4}=\frac{\rho \eta}{2},~~~~~{L}^1_{~4}=\eta,~~~{L}^4_{~1}=-\frac{\rho^2\eta}{4},~~~~~{L}^3_{~1}={L}^4_{~2}=\alpha,~~~~{L}^3_{~2}=\beta,~~~~{L}^1_{~2}=-{L}^3_{~4}=\frac{2\alpha}{\rho}$    \\\hline
       \end{tabular}}}
\end{center}
$\\$
Now to obtain the integrable sigma model we must consider the ad-invariant metric on the $h_{4}$.
An ad-invariant metric  $\Omega_{ij}$ so that it satisfies the condition \cite{Witten}
\begin{eqnarray}\label{5.1}
 f^{l}_{~ij} \;\Omega_{lk}+  f^{l}_{~ik}\;\Omega_{lj}\;=\;0,
\end{eqnarray}
can be obtained as follows \cite{EP}
\begin{eqnarray}\label{5.2}
	\Omega_{ij}=\left( \begin{tabular}{cccc}
                 $\rho$ & 0 & 0 & -$1$ \\
                 0 & 0 & $1$ & 0 \\
                 0 & $1$& 0 & 0 \\
                 -$1$ & 0 & 0 & 0 \\
                 \end{tabular} \right),
	\end{eqnarray}
for some real constant $\rho$.
In order to calculate the left-invariant one-forms we parameterize the $H_{4}$ group manifold
with the coordinates $x^{\mu} =(x, y, u, v)$. Thus, using the following representation for elements of the $H_{4}$
\begin{eqnarray}\label{5.3}
g = e^{v T_{4} } ~ e^{u T_{3}} ~ e^{x T_{1}}~e^{y T_{2}},
\end{eqnarray}
and then using the definition $g^{-1}d g={(g^{-1}d g)}^{i}T_{i}$ one can obtain the corresponding left-invariant one-forms, giving us \cite{EP}
\begin{eqnarray}\label{5.4}
(g^{-1}d g)^{1}&=&d x, \nonumber\\
(g^{-1}d g)^{2}&=&y d x+d y,\nonumber\\
(g^{-1}d g)^{3}&=&  e^{x} d u,\nonumber\\
(g^{-1}d g)^{4}&=& y e^{x} d u +d v.
\end{eqnarray}
Finally, by using \eqref{4.7.1}, \eqref{5.2}, \eqref{5.3} and \eqref{5.4} together with the inequivalent Haantjes structures of  Table 2 one can obtain the integrable sigma models on the  $H_{4}$.
In this manner, by using the Haantjes structures $L^{(h_{4})}_{_{47}}, L^{(h_{4})}_{_{48}}$ and $L^{(h_{4})}_{_{49}}$ we find three integrable sigma models
so that the backgrounds including metric and B-field are represented in Table 3.

\begin{center}
\small {{{\bf Table 3.}~ Integrable sigma models with Haantjes structure on the $H_{4}$}}
		{\scriptsize
			\renewcommand{\arraystretch}{1.5}{
\begin{tabular}{|p{0.6cm}|l|} \hline \hline
 &~~~~~~~~~~~~~~~~~Backgrounds including metric and $B$-field  \\ \hline
{$L^{(h_{4})}_{_{47}}$}~~~ &  ~~~~
$ds^{2}=\rho(a_{1}\epsilon y+1) dx^2+ \rho a_{1}\epsilon  dydx -2 a_{1}\epsilon y^{2}e^{x}dxdu-2 (a_{1}\epsilon y+1)dxdv-2 e^{x}(a_{1}\epsilon y-1)dydu-2 a_{1}\epsilon dydv$\\
 &
~~~~~~~$B=-\frac{1}{2}a_{1}\rho \epsilon dx \wedge dy $ \\\hline
{$L^{(h_{4})}_{_{48}}$}~~~ &  ~~~~
$ds^{2}= [a_{1}((2\delta y+\xi)-\delta \rho)+\rho]dx^2+a_{1} e^{2x}
(2\delta y+\xi) du^{2}+2a_{1} \delta dxdy+a_{1} e^{x}(4 \delta y-\delta \rho-2\xi)dxdu$ \\
 &
$~~~~~~~~~~~~~~~+2(a_{1}\delta-1)dxdv+2 e^{x}(a_{1}\delta+1)dydu+2 a_{1} \delta e^{x}dudv$\\
 &
~~~~~~~$B=\frac{1}{2}a_{1}e^{x} (\rho \delta+2 \xi)   dx \wedge du $ \\\hline
{$L^{(h_{4})}_{_{49}}$}~~~ &  ~~~~
$ds^{2}= (\frac{4a_{1} y (\beta y+2\alpha)-a_{1}\eta \rho^{2}+4\rho}{4})dx^2+a_{1}\beta dy^{2}-a_{1} \eta y^{2} e^{2x} du^{2}-a_{1} \eta dv^{2}+2a_{1}(\beta y+\alpha) dxdy-\frac{4a_{1} y e^{x}(y \alpha-\frac{\eta \rho^{2}}{4})}{\rho}dxdu$ \\
 &
$~~~~~~~~~~~~~~-a_{1}(\frac{4\alpha y -\eta \rho^{2}+\frac{2\rho}{a_{1}}}{\rho})dxdv-\frac{4a_{1} e^{x}(\alpha y-\frac{\rho}{2a_{1}})}{\rho}dydu-\frac{4 a_{1} \alpha}{\rho}dydv-2 a_{1}\eta ye^{x}dudv$\\
 &
~~~~~~~$B=0 $ \\\hline

\end{tabular}}}
\end{center}
As mentioned in the Introduction section, the Yang-Baxter deformations of Wess-Zumino-Witten model on the ${H_{4}}$ Lie group
and their corresponding non-Abelian duals were examined in \cite{EP,egh.gh.rez}, respectively.
Compared with the results of \cite{EP,egh.gh.rez}, the sigma models presented in Table 3 are new integrable ones.

Before closing this section, let us discuss the relationship between Nijenhuis and Haantjes structures.
In fact, we shall investigate whether such structures are completely independent or possibly related.
This discussion leads us to answering the question of why we compare our new deformed models with Yang-Baxter deformed models of \cite{EP,egh.gh.rez}.
First, according to formula \eqref{2.1} the components of \eqref{2.7} on the group manifold can be written as
\begin{eqnarray}\label{5.5}
N^k_{~ij}={f}^k_{~lm}~{L}^l_{~i}~{L}^m_{~j}-{f}^m_{~il}~{L}^l_{~j}~{L}^k_{~m}-{f}^m_{~lj}~{L}^l_{~i}~{L}^k_{~m}+{f}^l_{~ij}~{L}^m_{~l}~{L}^k_{~m},
\end{eqnarray}
where $N^k_{~ij} = e_{\lambda}^{~k} e^{\mu}_{~i} e^{\nu}_{~j} ~ N^\lambda_{~ \mu \nu}$. When this Nijenhuis torsion is equivalent to the ${f'}^k_{~ij}$ which are the  
structure constants of Lie algebra isomorphic to the $h_4$ (with the isomorphism $C$, i.e., $C_i^{~m}~ {f}^k_{~mn}~ C_j^{~n} = {f'}^l_{~ij} ~ C_l^{~k}$), the vanishing of the
Haantjes torsion defined by \eqref{2.3} (or \eqref{2.7}) is equivalent to the vanishing of Nijenhuis torsion of \eqref{2.1} with the structure constants 
$N^k_{~ij} = {f'}^k_{~ij}$. In this case, the solution of equation \eqref{2.7} is equivalent to the solution of \eqref{2.1} with the ${f'}^k_{~ij}$.
Accordingly, we will not have new integrable models and the results are equivalent to the integrable sigma models of the Yang-Baxter type \cite{EP,egh.gh.rez}.


\section{Conclusions}

In this paper we have written the formulation of the Haantjes structure in the Lie algebra framework.
We have solved the algebraic relations for the conditions of
Haantjes structure, and then by using the automorphism group of the ${h_{4}}$ Lie algebra
we have shown that the corresponding algebraic Haantjes structures are split into 49 inequivalent families.
We proposed a new deformation of the chiral sigma model on a Lie group by using Haantjes structure on it.
Then, using the ${h_{4}}$ Haantjes structures and solving the conditions on the integrability of the proposed model,
we managed to find three new integrable sigma models on the ${H_{4}}$ Lie group. Finally,
by comparing our results with the Yang-Baxter deformed backgrounds of the ${H_{4}}$ Wess-Zumino-Witten model \cite{EP}
and their non-Abelian duals \cite{egh.gh.rez}, it was concluded that our models were different from theirs.

It would be interesting to generalize the formulation of the algebraic Haantjes structure to a
Lie superalgebra case. In this way, we will be able to build new integrable sigma models with the super Haantjes structure that are significant
in comparison to Yang-Baxter deformed Wess-Zumino-Witten models based on Lie supergroups in low dimensions \cite{superYB,gauged.WZW}.
(see, also, \cite{osp} for the case of the OSP$(1|2)$ Lie supergroup).
We intend to address this problem in the future.

\subsection*{Declaration of competing interest}

The authors declare that they have no known competing financial
interests or personal relationships that could have appeared to influence the work reported in this paper.


\subsection*{Acknowledgements}

This work has been supported by the research vice chancellor of Azarbaijan Shahid Madani University under research fund No. 1402/231.
The authors gratefully thank to the Referee for the constructive comments and recommendations
which definitely help to improve the readability and quality of the paper.

\subsection*{Data availability statement}

No data was used for the research described in the article.
\\
\\
{\bf ORCID iDs}
\\
Ali Eghbali ~  https://orcid.org/0000-0001-6076-2179
\\
Adel Rezaei-Aghdam ~ https://orcid.org/0000-0003-4754-7911


\end{document}